# Rutile GeO$_2$: an ultrawide-band-gap semiconductor with ambipolar doping

S. Chae, J. Lee, K. A. Mengle, J. T. Heron, and E. Kioupakis[a]

*Department of Materials Science and Engineering, University of Michigan, Ann Arbor, MI 48109, USA*

Ultra-wide-band-gap (UWBG) semiconductors have tremendous potential to advance electronic devices as device performance improves superlinearly with increasing gap. Ambipolar doping, however, has been a major challenge for UWBG materials as dopant ionization energy and charge compensation generally increase with increasing band gap and significantly limit the semiconductor devices that can currently be realized. Using hybrid density functional theory, we demonstrate rutile germanium oxide (r-GeO$_2$) to be an alternative UWBG (4.68 eV) material that can be ambipolarly doped. We identify Sb$_{Ge}$, As$_{Ge}$, and F$_O$ as possible donors with low ionization energies and propose growth conditions to avoid charge compensation by deep acceptors such as V$_{Ge}$ and N$_O$. On the other hand, acceptors such as Al$_{Ge}$ have relatively large ionization energies (0.45 eV) due to the formation of localized hole polarons and are likely to be passivated by V$_O$, Ge$_i$, and self-interstitials. Yet, we find that the co-incorporation of Al$_{Ge}$ with interstitial H can increase the solubility limit of Al and enable hole conduction in the impurity band. Our results show that r-GeO$_2$ is a promising UWBG semiconductor that can overcome current doping challenges and enable the next generation of power electronics devices.

---

[a] Author to whom correspondence should be addressed. Electronic mail: kioup@umich.edu

Wide-band-gap (WBG) semiconductors are becoming widely employed in many applications such as solid-state lighting, transparent electrodes, and power electronics. Since a wider band gap allows a material to tolerate stronger electric fields, ultrawide-band-gap (UWBG) semiconductors, with band gap larger than 3.4 eV, are particularly advantageous in high-power applications as well as deep-UV optoelectronics.[1] Thus, for continuous device improvement, great effort has been taken toward finding alternative UWBG materials with superior properties over the state of the art. However, to employ materials in electronic applications, achieving control over the majority carrier type through doping is a prerequisite.

Doping of WBG materials is challenging, however.[2-4] For example, dopants in WBG semiconductors are often passivated by native defects since their formation energy linearly decreases with variations of the Fermi energy due to dopant incorporation.[2] Doping asymmetry is also a serious problem for WBG semiconductors. It arises from the fact that most WBG semiconductors are characterized by either a low valence-band maximum or a high conduction-band minimum relative to the vacuum level, which leads to high dopant ionization energies.[3] The emerging UWBG materials, e.g. AlN/AlGaN, $\beta$-$Ga_2O_3$, diamond, and h-BN, are facing such doping obstacles. For example, no good acceptors for AlN and $\beta$-$Ga_2O_3$ have been realized so far. Though Mg is a possible acceptor in AlGaN, the hole mobility is suppressed by alloy-disorder scattering.[5-6] Furthermore, holes in $\beta$-$Ga_2O_3$ form localized polarons even in the absence of dopants because of the flat valence band.[7] Efficient doping is more challenging for diamond[1] and h-BN[8-9] since dopant levels are deep for both acceptors and donors.

Here, we introduce rutile germanium oxide (r-$GeO_2$) as a promising UWBG material that can be ambipolarly doped. Though it has been studied in the context of glasses[10-11] and gate-dielectric layers,[12-13] $GeO_2$ is a promising, yet unexplored UWBG material. Among its five different polymorphs, the rutile structure of $GeO_2$ [Fig. 1(a)] is thermodynamically stable up to ~1300 ºC[14], chemically resistant when exposed to solvents[15], mechanically strong, and dense[16]. It has a direct band gap of 4.68 eV at the Γ point [Fig. 1(b)] and its closed-packed and highly symmetric rutile structure may lead to efficient electron and heat conduction. Thus, it may be a promising candidate for electronic and optoelectronic applications. However, its doping properties and its potential for semiconducting applications have not been previously investigated.

We apply first-principles calculations based on density functional theory (DFT) to predict the n-type and p-type dopability of r-$GeO_2$. We investigate issues regarding carrier localization and charge compensation by passivating native defects and suggest growth conditions and co-doping methods that increase dopant solubility and minimize charge compensation. Our results show that substitutional Sb, As, and F are efficient donors, while co-doping Al with H and subsequent annealing is a promising method to incorporate and activate acceptors with ionization energies of 0.45 eV. The solubility of Al enabled by codoping with H can exceed the Mott critical density and enable p-type conduction through the formation of an impurity band. We therefore predict that ambipolar doping is possible in r-$GeO_2$.

Our calculations are based on hybrid density functional theory using the projector augmented wave (PAW) method and the Heyd-Scuseria-Ernzerhof (HSE06)[17] functional as implemented in the Vienna Ab initio Simulation Package (VASP)[18-19]. The GW-compatible pseudopotentials are employed for Ge and O with a plane-wave cutoff energy of 400 eV. Structures were relaxed using the quasi-Newton algorithm with a maximal force criterion of 0.01eV/Å. To correct for the underestimated band gap, the amount of Hartree-Fock exchange was set to 35%. Using these parameters, our calculated values for the band gap (4.64 eV), lattice parameters (a=4.394 Å and c=2.866 Å), and enthalpy of formation (-5.49 eV/formula unit) are in excellent agreement with experiment (4.68 eV[20], a=4.398 Å, c=2.863 Å[21], and -5.62 eV/formula unit,[22] respectively). We evaluated the formation energies of neutral and charged defects as a function of the Fermi level and growth conditions using the methodology described in Ref. [23] We modeled point defects and impurities using 72-atom supercells and a 2×2×2 Γ-centered Brillouin-zone sampling mesh. A comparison of the formation energy between Γ-centered and Monkhorst-Pack sampling grids is shown in Fig. S1. We tested the convergence with respect to supercell size by comparing results for a 72- and a 162-atom cell for the O vacancy. The results show that the defect-formation energy and charge-transition level change by less than 0.12 eV and 0.06 eV, respectively. The correction energy due to the image charges and pseudopotential alignment introduced by the supercell approximation was calculated with the SXDEFECTALIGN code.[24] We set the static dielectric constant of the host material needed for the correction to $\varepsilon_0 = 13.28$, which is the directionally averaged value determined by density functional perturbation theory[25] and the Quantum ESPRESSO code,[26] and which closely matches the average (13.73) of the experimental values[27] ($\varepsilon_{0\parallel} = 14.5$ and $\varepsilon_{0\perp} = 12.2$). The chemical potentials vary depending on experimental conditions. The chemical potentials of Ge and O areset by the equation $\mu_{Ge} + 2\mu_O = \Delta H_f(GeO_2)$, where $\Delta H_f(GeO_2)$ is our calculated enthalpy of formation of r-$GeO_2$. Our defect calculations were performed at the two limits of extreme O-rich/Ge-poor [$\mu_O = 0$ eV and $\mu_{Ge} = \Delta H_f(GeO_2)$] and Ge-rich/O-poor [$\mu_{Ge} = 0$ and $\mu_O = \Delta H_f(GeO_2)$] environments. The chemical potentials of the impurity species are also limited by the formation of secondary phases such as $H_2O$, $Ge_3N_4$, $Sb_2O_3$, $As_2O_5$, $Bi_2O_5$, $GeF_4$, $Al_2O_3$, $In_2O_3$ and $Ga_2O_3$. The reference elemental phases are molecular $O_2$, $N_2$, $F_2$ and $H_2$ and the bulk phases of Ge, Sb, As, Bi, Al, Ga and In. Spin-polarized calculations were performed for defect supercells with odd numbers of electrons. Spin-orbit coupling effects were considered for calculations including the heavier Bi element.

We first explored the possibility of n-type doping of r-$GeO_2$ with Sb, As, Bi, and F dopants. Sb, As, and Bi are group-15 elements with ionic radius close to that of $Ge^{+4}$ and are expected to preferentially occupy the Ge-substitutional site. On the other hand, F is expected to replace an O atom, generating one extra electron. Fig. 2 shows that all of the investigated dopants, with the exception of Bi, are shallow donors, indicating promising n-type dopability of r-$GeO_2$. The formation energies of donors vary depending on the simulated growth conditions: $F_O$ forms more easily under O-poor conditions since F substitutes the O atom, while $Sb_{Ge}$ is the most promising donor under O-rich conditions. On the other hand, Bi is not a suitable donor for $GeO_2$ not only because it has a relatively high formation energy, but being a *p* element, it is also stable in the +3 valence state for Fermi levels above 3.81 eV and thus consumes free electrons. We also note that, unlike $SnO_2$ in which Sb is partially stable in the $Sb^{+3}$ valence state,[28] Sb in r-$GeO_2$ is only stable in the $Sb^{+5}$ state for the entire Fermi level range, thus 100% doping efficiency is expected. We also expect a higher solubility of both Sb and As in r-$GeO_2$ compared to $SnO_2$ owing to their lower formation energy.[29] Though Sb and As can be incorporated as interstitial dopants for Fermi energies near the valence band maximum (VBM), substitutional incorporation dominates for n-doped r-$GeO_2$.

We also investigated group-13 elements, i.e. Al, Ga, and In, to explore the p-type dopability of r-$GeO_2$. In order to investigate the formation of localized hole polarons (which inhibits p-type doping in competitor $\beta$-$Ga_2O_3$), the structure optimization for p-dopants was performed by intentionally displacing one oxygen atom from its symmetric position. We found that in the neutral charge state, all the p-type dopants studied here form a hole polaron, which is localized on an oxygen atom next to the dopant and is accompanied by a local distortion of the crystal lattice. We also found that Al and Ga dopants can trap an extra hole on the opposite side of the first one, further distorting the local bonds. As a result, the 0/-1 ionization energy was calculated to be 0.45 eV for $Al_{Ge}$, 0.54 eV for $Ga_{Ge}$ and 0.48 eV for $In_{Ge}$ (Table I). As shown in Fig. 3(a-b), $Al_{Ge}$ has low formation energy and lower ionization energy than $Al_{Sn}$ and $Ga_{Sn}$ in $SnO_2$ (1.15 eV and 1.05 eV, respectively),[30] and is thus predicted to be a possible (albeit not very efficient) p-type dopant candidate in r-$GeO_2$.

We further investigated the formation of hole polarons and the ionization of $Al_{Ge}$ using the configuration coordinate diagram in Fig. 3(c), which shows two atomic configurations of $Al_{Ge}$; one for delocalized and the other for localized holes.[31] Our calculations show that the localized hole in $Al_{Ge}$ is more stable than the delocalized one by 0.45 eV. The energy difference is the self-trapping energy of the polaron, $E_{ST}$, which also corresponds to the 0/-1 ionization energy of $Al_{Ge}$. The competing energy for the formation of a hole polaron is the strain energy for the lattice distortion, $E_S$. It can be calculated from the energy difference between the energy of the atomic configuration corresponding to the delocalized hole in the charge-neutral state and the energy of the atomic configuration corresponding to the localized hole, the result of which is 1.19 eV. Finally, the energy of 1.64 eV, the sum of the self-trapping energy ($E_{ST}$) and the lattice energy cost ($E_S$), is the energy required for the vertical transition of the hole polaron, which also represents the absorption energy in $Al_{Ge}$.

Next, we investigated native defects and common impurities in r-GeO$_2$ to identify possible sources of unintentional dopants and dopant compensation. We investigated the O vacancy ($V_O$), Ge vacancy ($V_{Ge}$), O interstitial ($O_i$), Ge interstitial ($Ge_i$), Ge on O antisite ($Ge_O$) intrinsic defects, and common impurities related to H, N, and C atoms since they are commonly present in the synthesis environment and can often be inadvertently incorporated into a material. Fig. 4 shows their formation energy as a function of the Fermi energy at the two extreme growth conditions.

We predict that possible sources of donor compensation in r-GeO$_2$ are $V_{Ge}$ in Ge-poor conditions and $N_O$ in O-poor conditions. $V_{Ge}$ is a shallow acceptor and has lower formation energy than donors studied here for Fermi energies near the conduction band minimum (CBM) in Ge-poor conditions. On the other hand, $N_O$ is likely to form under O-poor conditions and has deep-acceptor-like properties with an acceptor ionization energy of 3.03 eV. Thus, special care in choosing growth conditions and eliminating nitrogen during growth may be required to avoid charge compensation and enhance the doping efficiency of r-GeO$_2$.

We also note that native defects or impurities can act as a potential charge-compensation source for p-type dopants. $V_O$ may be the major passivating defect, which is a deep donor-type defect having lower formation energy in the low Fermi-energy region compared to any of the p-dopants studied here. $Ge_i$ and Ge$_O$ are also donor-type intrinsic defects, having low formation energies for Fermi energies near the VBM under Ge-rich conditions. Hydrogen-related defects are shallow donors, being stable in the +1 charge state in the entire Fermi-energy range within the gap. In addition, for Fermi levels near the VBM, Al, Ga, and In dopants prefer to incorporate into the interstitial site of r-GeO$_2$ by donating 3 electrons, resulting in self-passivation.

To achieve p-doping of r-GeO$_2$, it is crucial to avoid compensating defects and increase the solubility of acceptors. One strategy is co-doping acceptors with highly mobile hydrogen shallow donors and post-annealing in a reducing environment to activate holes. We thus investigated the effect of interstitial hydrogen on p-type doping of r-GeO$_2$ by calculating the formation energies of H$_i$-acceptor defect complexes. Our results in Table I and Fig. 5 show that H$_i$-acceptor complexes are stable in the neutral charge state, except for H-Al$_{Ge}$ at Fermi energies near the VBM. As shown in Fig. 5(b), H$_i$ primarily bonds with O while its atomic position moves toward the direction of the acceptor. The low formation energies of H$_i$-acceptor complexes, together with their atomic configurations, indicate their strong interaction, thus H co-doping can effectively enhance the solubility of acceptors. In order to reactivate the hole carrier, H needs to be dissociated. We thus determine the binding energy of the H-acceptor complex by comparing its formation energy to the isolated defects. Although the binding energies of 0.96 eV, 0.98 eV, and 0.92 eV for Al-H, Ga-H, and In-H, respectively, are high, dissociation is achievable using high-temperature post-annealing techniques such as rapid thermal annealing (RTA). This technique has been widely used for the p-type doping of GaN with Mg, in which thermal annealing at 700 °C[32] effectviely dissociates the H that binds to Mg with a binding energy of 0.7 eV.[33]

The large electron affinity of wide-band-gap oxides, as well as the formation of localized hole polarons, have been fundamental challenges for p-doping of oxide semiconductors in general. However, despite the ultra-wide band gap of r-GeO$_2$, our calculations predict possible p-type dopability of r-GeO$_2$ with Al acceptors. Though the 0.45 eV acceptor ionization energy is high, interactions between acceptors broaden the acceptor band and lower the effective ionization energy. Moreover, hole conduction is enabled by impurity-band formation for acceptor concentrations exceeding the Mott-transition limit of approximately $(0.2/a_H)^3$ where $a_H$ is the acceptor-bound hole wave function radius.[34] In r-GeO$_2$, we estimate $a_H$ to be 3.38 Å and the critical Mott density for Al acceptors to be $2.07 \times 10^{20}$ cm$^{-3}$. Considering the low formation energy of the Al-H complex, this concentration can be easily achieved at a growth temperature above 536 ºC. Thus, we expect that for heavily p-typed doped r-GeO$_2$, the activation energy becomes lower than the isolated-acceptor ionization energy and enables p-type conduction through the impurity band. Whereas AlN and $\beta$-Ga$_2$O$_3$ have large hole effective masses ($m_h^* \sim 7.26$ for AlN [35] and $m_h^* \sim 40$ for $\beta$-Ga$_2$O$_3$ [36]) that make hole conduction very challenging, r-GeO$_2$ has much smaller hole effective mass ($m_h^\perp = 1.28$, $m_h^\parallel = 1.74$) leading to a more delocalized hole wave function and smaller Mott transition concentration, which can be easily achieved by co-doping with H$_i$. Also, despite its wider band gap, the p-type doping of r-GeO$_2$ is more promising compared to SnO$_2$ or ZnO as a consequence of its smaller effective mass and larger dielectric constant. A comparison of the acceptor ionization energies and valence band offsets between r-GeO$_2$, SnO$_2$, and ZnO is shown in Table S1 and Fig. S2. The uniqueness of rutile GeO$_2$ that enables p-type conduction compared to other UWBG semiconductors and n-type WBG oxide semiconductors originates from its small hole effective mass, large dielectric constant, and the strong hydrogen-acceptor interaction that effectively lowers the formation energy of acceptors.

In conclusion, we investigated the formation of point defects and the n-/p-type dopability of r-GeO$_2$. We found that efficient n-type doping of r-GeO$_2$ can be achieved with Sb, As, and F dopants under appropriate growth conditions. We also suggest the possibility of p-type doping of r-GeO$_2$ with Al dopants with an ionization energy of 0.45 eV. Co-doping acceptors with hydrogen and subsequent annealing can overcome the passivation from compensating native defects and reach acceptor concentrations that enable hole conduction through an impurity band.

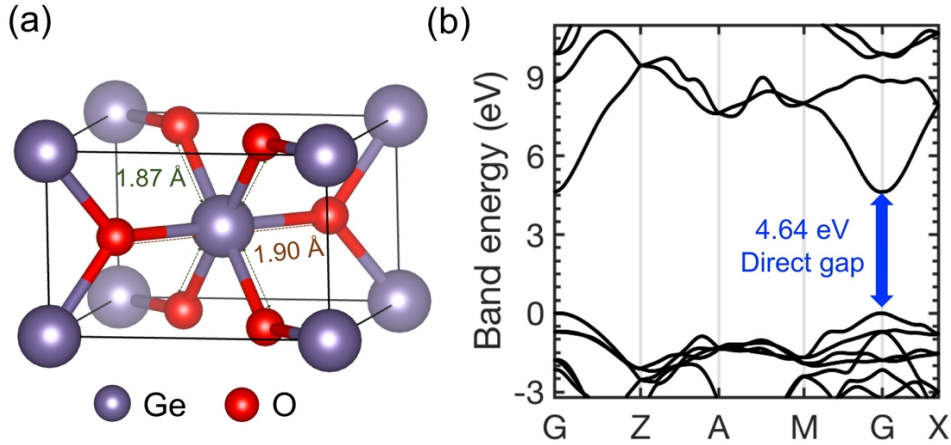

FIG. 1. (a) Crystal structure of rutile $GeO_2$. (b) Our calculated band structure of rutile $GeO_2$ using the HSE06 hybrid functional.

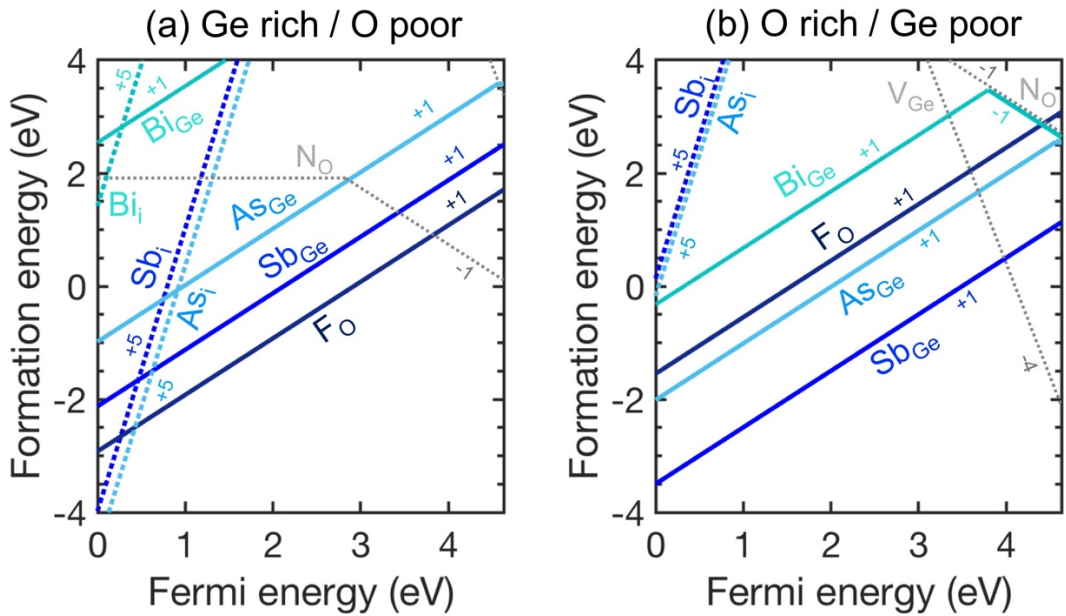

FIG. 2. Formation energy of donor defects and potential charge-compensating native defects as a function of the Fermi level in the limit of (a) Ge rich / O poor and (b) O rich / Ge poor conditions.

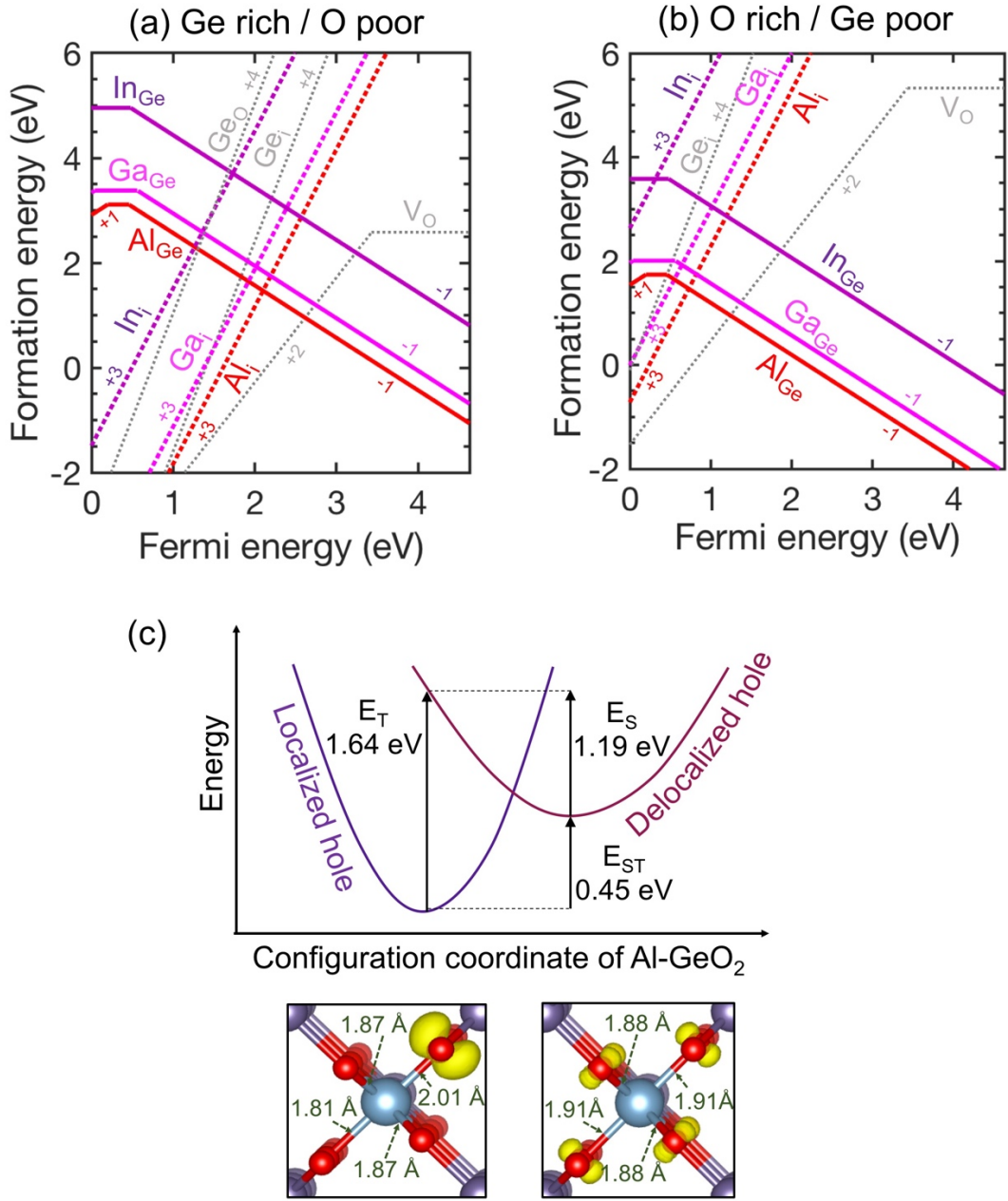

FIG. 3. (a-b) Formation energy of acceptor defects and potential charge-compensating intrinsic defects as a function of the Fermi level in the limit of (a) Ge rich / O poor and (b) O rich / Ge poor conditions. (c) Configuration coordinate diagram for the formation of localized hole polarons in Al-doped r-GeO$_2$. E$_T$, E$_{ST}$ and E$_S$ indicate the vertical excitation energy, the polaron self-trapping energy, and the strain energy, respectively. The insets show the isosurface of the band-decomposed charge density at the VBM for the localized and the delocalized holes near an Al$_{Ge}$ dopant.

TABLE I. Acceptor-type dopants in r-GeO$_2$, their ionization energies, and their binding energies with hydrogen interstitials.

| Acceptor | Ionization energy (eV) | $E_b$ for H dissociation (eV) |
|---|---|---|
| $Al_{Ge}$ | 0.45 | 0.96 |
| $Ga_{Ge}$ | 0.54 | 0.98 |
| $In_{Ge}$ | 0.48 | 0.92 |

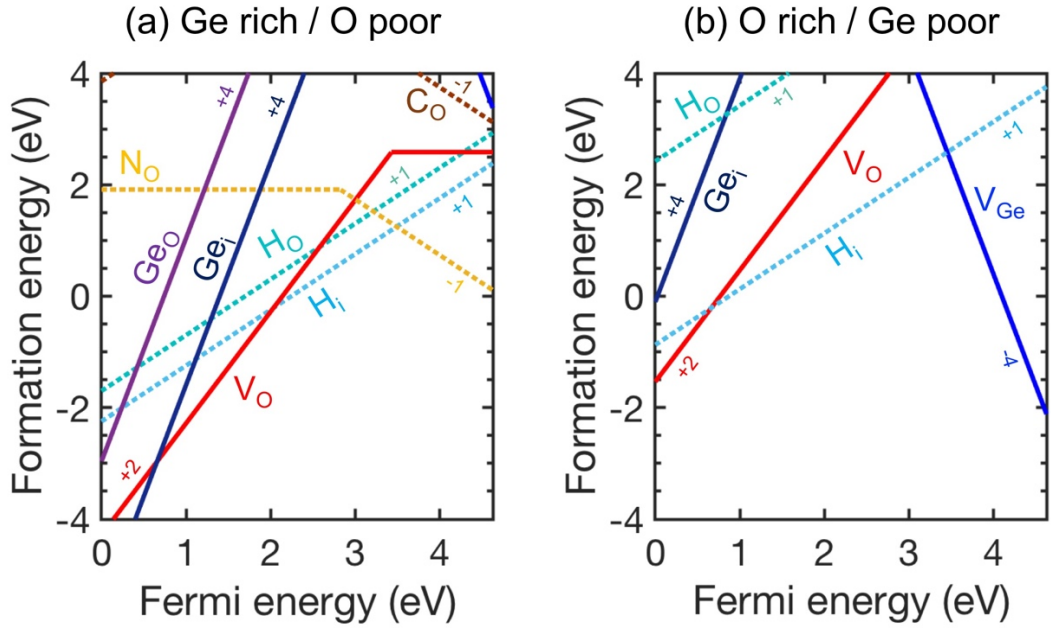

FIG. 4. Formation energy of intrinsic point defects and common impurities as a function of the Fermi level (referenced to the valence-band maximum) in the limit of (a) Ge rich / O poor and (b) O rich / Ge poor conditions.

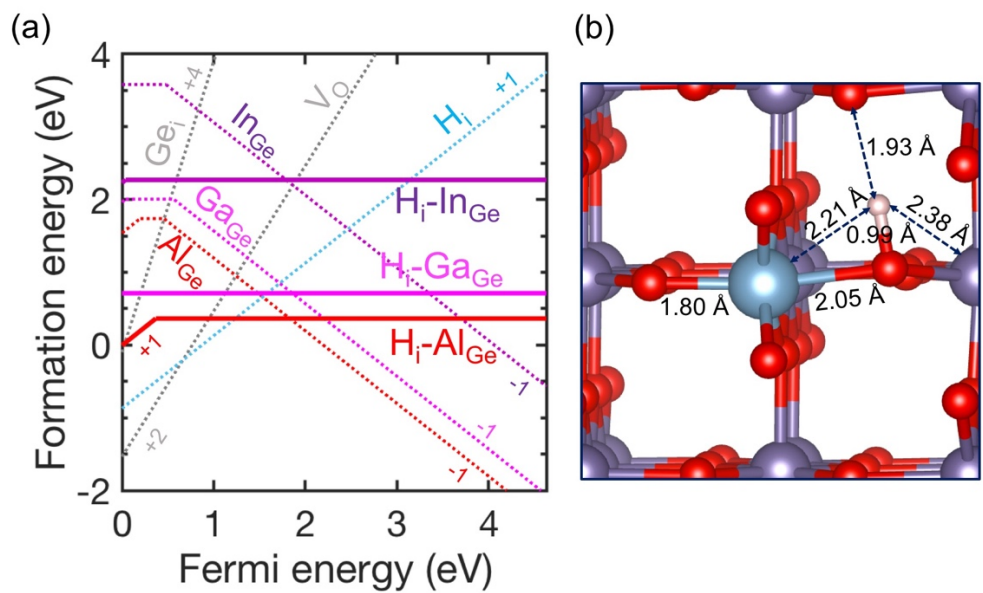

FIG. 5. (a) Formation energy of $H_i$-acceptor defect complexes as a function of the Fermi level plotted in comparison with $H_i$, acceptors, and intrinsic donor-type defects in the limit of Ge poor / O rich conditions. (b) The atomic configuration of the $H_i$-$Al_{Ge}$ defect complex in r-GeO$_2$.

## SUPPLEMENTARY MATERIAL

See supplementary material for the comparison of Γ-centered and Monkhorst-Pack Brilliouin-zone sampling grids to calculate the formation energy of oxygen vacancy in r-GeO$_2$, the acceptor ionization energies and the band offsets of ZnO, SnO$_2$, and r-GeO$_2$, and the formation energy of $V_O$, $V_{Ge}$, and $H_i$-$Al_{Ge}$ in r-GeO$_2$ as a function of Fermi energy.

## ACKNOWLEDGMENTS

This work was supported by the Designing Materials to Revolutionize and Engineer our Future (DMREF) Program under Award No. 1534221, funded by the National Science Foundation. K.A.M. acknowledges the support from the National Science Foundation Graduate Research Fellowship Program through Grant No. DGE 1256260. This research used resources of the National Energy

Research Scientific Computing Center, a DOE office of Science User Facility supported by the Office of Science of the U.S. Department of Energy under Contract No. DE-AC02-05CH11231.

# Supplementary Material

# Rutile GeO$_2$: an ultrawide-band-gap semiconductor with ambipolar doping

S. Chae, J. Lee, K. A. Mengle, J. T. Heron, and E. Kioupakis[1]

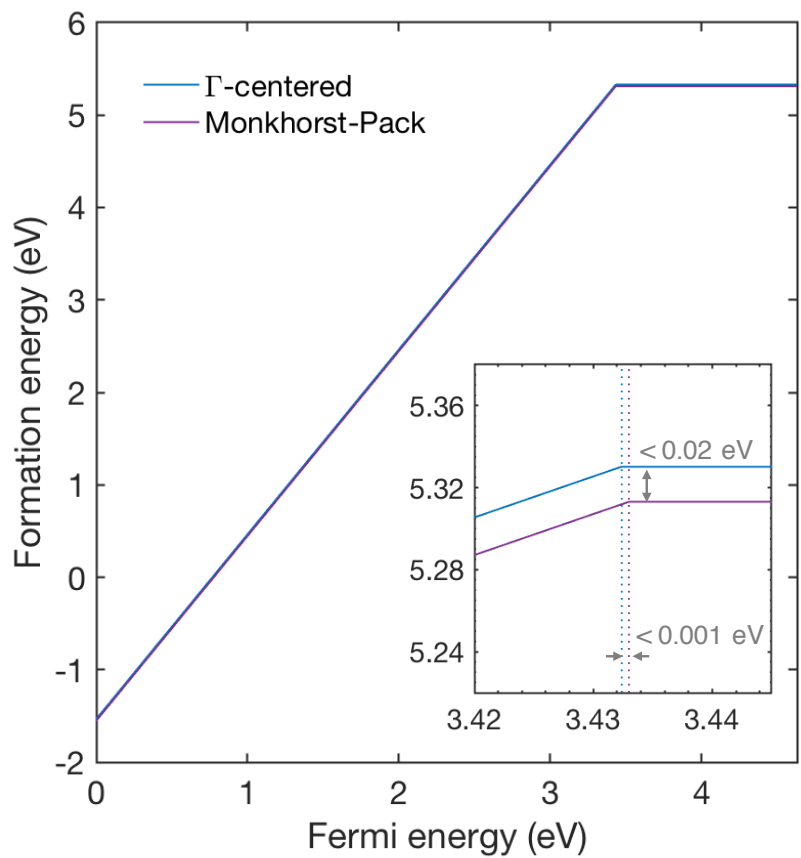

FIG. S1. Comparison of formation energy of oxygen vacancy in r-GeO$_2$ calculated by using Γ-centered and Monkhorst-Pack Brillouin-zone sampling grids. O rich / Ge poor conditions are used for the chemical potentials.

[1] Author to whom correspondence should be addressed. Electronic mail: kioup@umich.edu

Table S1. Acceptor ionization energies of ZnO, SnO$_2$, and r-GeO$_2$. The values for ZnO and SnO$_2$ are taken from ref [S1]. r-GeO$_2$ is predicted to have shallow acceptor ionization energies compared to ZnO and SnO$_2$, which may originate from the smaller hole effective mass of r-GeO$_2$ (m$_h$* = 1.44 for GeO$_2$, 2.28[S2] for ZnO, and 1.64 for SnO$_2$[S2]) and its larger dielectric constant ($\varepsilon$ = 13.73 for r-GeO$_2$, 12.62 for SnO$_2$,[S3] and 8.53 for ZnO[S4]).

| ZnO | Li | Na | Ag |
|---|---|---|---|
| $\varepsilon_A$ (0/-) | 0.86 | 0.79 | 1.18 |

| SnO$_2$ | Al | Ga | In |
|---|---|---|---|
| $\varepsilon_A$ (0/-) | 0.85 | 0.76 | 0.58 |

| r-GeO$_2$ | Al | Ga | In |
|---|---|---|---|
| $\varepsilon_A$ (0/-) | 0.45 | 0.54 | 0.48 |

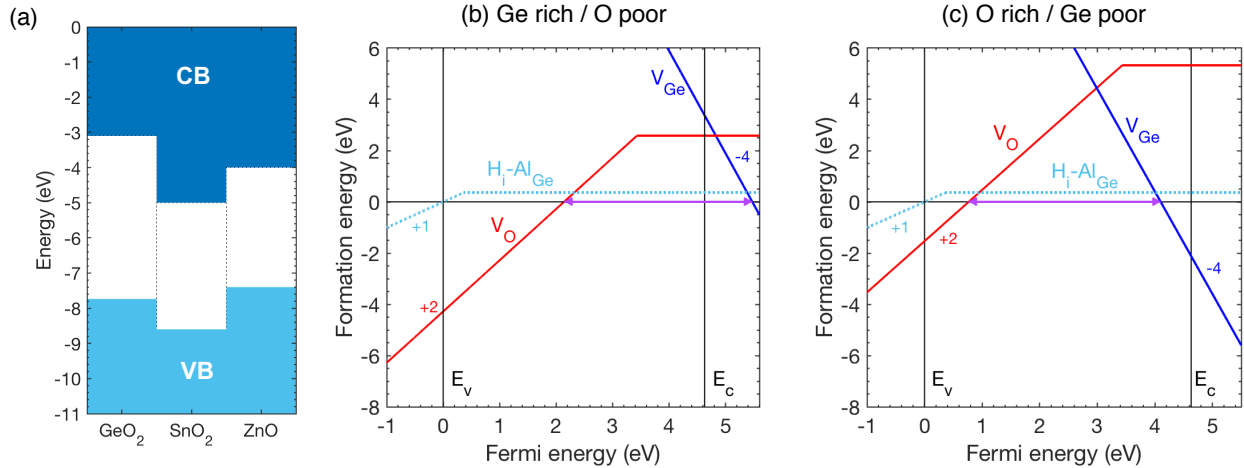

FIG. S2. (a) The band offsets between r-GeO$_2$ (our calculation), SnO$_2$, and ZnO (values taken from ref [S5]). (b-c) The formation energy of V$_O$, V$_{Ge}$, and H$_i$-Al$_{Ge}$ as a function of Fermi energy for (b) extremely Ge rich / O poor and (c) O rich / Ge poor conditions.